\documentclass[conference,compsoc]{IEEEtran}

\makeatletter
\newcommand\footnoteref[1]{\protected@xdef\@thefnmark{\ref{#1}}\@footnotemark}
\makeatother

\ifCLASSOPTIONcompsoc
  \usepackage[nocompress]{cite}
\else
  \usepackage{cite}
\fi

\usepackage[pdftex]{graphicx}
\DeclareGraphicsExtensions{.pdf,.jpeg,.png}

\usepackage[caption=false,font=footnotesize,labelfont=sf,textfont=sf]{subfig}

\usepackage{amsmath}
\usepackage{amssymb}
\usepackage{pifont}

\usepackage{algorithmic}

\usepackage{array}

\usepackage{url}

\usepackage{enumitem}
\usepackage{parcolumns}
\usepackage{listings}

\lstdefinelanguage
   [x64]{Assembler}     
   [x86masm]{Assembler} 
   {morekeywords={CDQE,CQO,CMPSQ,CMPXCHG16B,JRCXZ,LODSQ,MOVSXD, %
                  POPFQ,PUSHFQ,SCASQ,STOSQ,IRETQ,RDTSCP,SWAPGS,retq} %
                  } 

\usepackage{textcomp}

\usepackage{xcolor}

\begin{document}

\IEEEoverridecommandlockouts

\title{DynSGX: A Privacy Preserving Toolset for Dynamically
Loading\\Functions into Intel(R) SGX Enclaves}
\author{
\IEEEauthorblockN{Rodolfo Silva, Pedro Barbosa, Andrey Brito}
\IEEEauthorblockA{Universidade Federal de Campina Grande\\
Campina Grande, PB, Brazil \\
 \{rodolfoams, pedro, andrey\}@lsd.ufcg.edu.br}
}

%

\IEEEpubid{\begin{minipage}{\textwidth}\ \\[12pt]
    \copyright~\copyright~2017 IEEE. Personal use of this material is permitted. Permission from IEEE\\must be obtained for all other uses, in any current or future media, including\\reprinting/republishing this material for advertising or promotional purposes,\\creating new collective works, for resale or redistribution to servers or lists, or reuse\\of any copyrighted component of this work in other works
\end{minipage}}

\maketitle

\begin{abstract}
	Intel(R) Software Guard eXtensions (SGX) is a hardware-based technology for
	ensuring security of sensitive data from disclosure or modification that
	enables user-level applications to allocate protected areas of memory called
    enclaves. Such memory areas are cryptographically protected even from code
    running	with higher privilege levels. This memory protection can be used to
    develop secure and dependable applications, but the technology has some limitations: ($i$) the code of an enclave is
     visible at load time, ($ii$) libraries used by the code must be
    statically linked, and ($iii$) the protected memory size is limited,
    demanding page swapping to be done when this limit is exceeded. We present
    DynSGX, a privacy preserving tool that enables users
	and developers to dynamically load and unload code to be executed inside SGX
	enclaves. Such a technology makes possible that developers use public cloud infrastructures to run applications based on sensitive code and data. Moreover, we present a series of experiments that assess how applications dynamically loaded by DynSGX perform in comparison to statically linked applications that
    disregard privacy of the enclave code at load time.
\end{abstract}


\section{INTRODUCTION}
We live in an increasingly more connected world, where people are constantly uploading
personal data to environments that cannot be controlled by them. Sometimes this
data is meant to become public, but in most cases they are 
concerns about its confidentiality and/or integrity.
In that sense, developers need to find ways to
protect their users' data from theft or improper modification at all costs.

Besides that, developers and companies often host their applications in public
cloud environments in order to increase their availability and scalability, or
even to lower costs spent on physical infrastructure~\cite{armbrust2010view}. In
such environments, multiple applications from different owners may reside in the
same physical server, making it possible for rogue cloud users to exploit
existing security breaches to obtain secret data from other users.


Security of data based solely on software usually falls short due to
vulnerabilities existing in the application developed, in libraries/resources
used by the application, or even in the operating system/virtualization technology~\cite{gkortzis2016empirical}. Considering that, higher security levels have been
demanded.


In response to this demand, over the course of the past decade, several efforts were made to define and
implement a security enabler called Trusted Execution Environment (TEE)~\cite{omtp2009, logictrusted, mccune2008flicker}. To put in a few words, TEE is a secure area of the
main processor, or a separate, dedicated processor, that guarantees that code and data stored inside it will not be
modified or disclosed without permission. It was first specified by
GlobalPlatform~\cite{platform2011trusted}, and since then has attracted
attention from several companies that have provided their own implementations~\cite{ARMTrustZone, mckeen2013innovative, amdSMESEV}.

One of these implementations is Intel SGX~\cite{mckeen2013innovative}. It is available on recent off-the-shelf processors
based on the Skylake
microarchitecture or newer, and already has
a wide variety of research publications related to its applicability on real
world scenarios~\cite{hoekstra2013using}.
This technology allows user-level code to be executed inside
protected areas of memory, named enclaves. This technology also provides means for
users/applications to verify if a given application is running inside an Intel SGX
enclave, and even check if the code of the application is indeed the one expected to
be running there, a process called Remote Attestation~\cite{anati2013innovative}. 

Despite these advantages, Intel SGX has some
limitations regarding memory usage and code privacy that may raise some concerns
when using it in cloud environments. Regarding memory, only a very limited
area, up to $128\ MB$ in size, can be protected by the processor\footnote{\url{https://download.01.org/intel-sgx/linux-1.9/docs/Intel_SGX_SDK_Developer_Reference_Linux_1.9_Open_Source.pdf}}. When this limit is reached, data need to be swapped to/from the
unprotected DRAM, generating an overhead. Concerning code privacy, SGX
programming model does not prevent code disclosure, since all enclave code can be viewed from the executable file stored in the file system. These limitations will 
be better discussed in the next sections of this paper.

In order to overcome these limitations, we propose DynSGX, a 
privacy preserving toolset for dynamically loading functions into and unloading
functions from Intel SGX enclaves. Our toolset enables developers to better manage the
scarce memory resources they have available to use with Intel SGX, as well as to
keep their applications private even when being loaded into cloud environments.

The rest of the paper is divided as follows: in Section~\ref{sec:intelsgx} we
get into more details about Intel SGX, its limitations and possible
vulnerabilities. We continue by introducing DynSGX in Section~\ref{sec:dynsgx}.
Further on, in Section~\ref{sec:evaluation}, we present an evaluation of our
proposed solution. In Section~\ref{sec:relatedwork} we discuss published work related to our toolset.
Finally, in Section~\ref{sec:conclusion} we draw our
conclusions, and describe some possible future work to further improve our
toolset.

\section{INTEL SGX}
\label{sec:intelsgx}

Intel Software Guard eXtensions (SGX) can be described as a new set of
instructions and changes in memory access mechanisms added to the Intel
Architecture. It is a hardware-based technology that, like other TEE
implementations, is used for ensuring security of sensitive data from disclosure
or modification~\cite{mckeen2013innovative}. 
SGX works as an ``inverse sandbox'' mechanism, where code can be sealed inside an enclave (i.e., private region of
memory). Inside the enclave, code and data are protected  by  hardware
enforced  access  control  policies which prevent attacks against the  enclave's
content even when these originate from privileged software such as virtual
machine monitors and operating systems.

The protection of the enclaves is ensured by the creation of a reserved area of
memory called Processor Reserved Memory (PRM), which is reserved by BIOS at boot
time. Inside the PRM lies another region of memory known as Enclave Page Cache
(EPC), also configured by BIOS at boot time, where the enclaves' pages actually
reside. Access to these pages is controlled by the processor and protected by
mechanisms such as the Memory Encryption Engine (MEE).

SGX also allows users to perform a Remote Attestation (RA) process (i.e.
cryptographically verify if the desired application is running inside an SGX
enclave).  The generation of the hardware-based material used in the RA process
is also enabled by the SGX instructions~\cite{anati2013innovative}.
The RA process can be used to establish a secure communication channel between
an application enclave and a user, by sharing a symmetric key via an Elliptic
Curve Diffie-Hellman (ECDH) protocol. Users can then use this key to securely
exchange messages with an application running inside an enclave.

Possible applications of SGX have been discussed in
\cite{hoekstra2013using}, where  examples of applications that make use of the
SGX capabilities were presented, as well as an application architecture
including an application split between components requiring security protection
which should run within enclaves, and components that do not require protection
and can therefore be executed outside enclaves. In~\cite{silva2017}, SGX is used
for securely communicating and aggregating fine-grained smart metering data in a
cloud environment.

\subsection{SGX Components}

The SGX solution comprises four main components: ($i$) the set of
instructions in the processor, ($ii$) the operating system drivers, ($iii$) the
Software Development Kit (SDK), and ($iv$) the Platform Software (PSW).

\subsubsection{SGX Instructions Set}
SGX instructions set is available on off-the-shelf processors based on the
Skylake microarchitecture or newer, starting from the 6\textsuperscript{th}
Generation Intel Core family and on the 5\textsuperscript{th} Xeon processors. It consists of 17
new instructions that can be classified into the following functions
\cite{mckeen2013innovative}:

\begin{itemize}
\item \textbf{Enclave build/teardown}: Used to allocate protected memory for the
enclave, load values into the protected memory, measure the values loaded into
the enclave's protected memory, and tear down the enclave after the application
has finished.

\item \textbf{Enclave entry/exit}: Used to enter and exit the enclave. An
enclave can be entered and exited explicitly. It may also be exited
asynchronously due to interrupts or exceptions. In the case of asynchronous
exits, the hardware will save all secrets inside the enclave, scrub secrets from
registers, and return to external program flow. Later, it then resumes where it left
off execution.

\item \textbf{Enclave security operations}: Enable an enclave to prove to an
external party that the enclave was built on hardware which supports the SGX
instruction set.

\item \textbf{Paging instructions}: Allow system software to securely move
enclave pages to and from unprotected memory.

\item \textbf{Debug instructions}: Allow developers to use familiar	debugging
techniques inside special debug enclaves. A debug enclave can be single stepped
and examined. A debug enclave cannot share data with a production enclave. This
protects enclave developers if a debug enclave should escape the development
environment.
\end{itemize}

\subsubsection{SGX Drivers}
The SGX \emph{drivers} enable OSs and other softwares to access the SGX
hardware. Intel SGX drivers are available both for Windows (via Intel Management
Engine)\footnote{\label{fn:SGXSDK}\url{https://software.intel.com/en-us/sgx-sdk/download}} and for Linux*\footnote{\url{https://github.com/01org/linux-sgx-driver}} platforms.
They serve as an abstraction to enable developers to
write higher-level code for using the device capabilities.

\subsubsection{SGX SDK}
The SGX Software Development Kit (SDK) is a collection of APIs, sample source
code, libraries and tools that enable software developers to write and debug SGX
applications in C/C++. Intel SGX SDK is available both for Windows, and for
Linux* platforms\footnoteref{fn:SGXSDK}.

\subsubsection{SGX PSW}
The SGX \textit{Platform Software} (PSW) is a collection of special SGX
enclaves, and an Intel SGX Application Enclave Services Manager (AESM),
provided along with the SGX SDK. These special enclaves and AESM are used when
loading enclaves, retrieving cryptographic keys, and evaluating the contents of
an enclave.

\subsection{SGX Memory Management}

The memory management model of Intel SGX makes it very useful for providing
data security. The main aspects of the memory model are discussed below:

\begin{itemize}
	\item \textbf{Enclave Page Cache}:
	The Enclave Page Cache (EPC) is protected memory used to store
	enclave pages and SGX structures. The EPC is divided into 4KB chunks called
	EPC pages. EPC pages can either be valid or invalid. A valid EPC page
	contains either an enclave page or an SGX structure.
	Each enclave instance has an enclave control structure, SECS. Every valid
	enclave page in the EPC belongs to exactly one enclave instance. System
	software is required to map enclave virtual addresses to a valid EPC page.

	\item \textbf{Memory Encryption Engine}:
	Memory Encryption Engine (MEE) is a hardware unit that encrypts and
	protects the integrity of selected traffic between the processor package and the
	main memory (DRAM). The overall memory region that an MEE operates on is
    called an MEE Region. Depending on implementation, the PRM is covered by
	one or more MEE regions.

	Intel SGX guarantees that all the data that leaves the CPU and is stored in
	DRAM is first encrypted using the MEE. Thus, even attackers with physical
	access to DRAM will not be able to retrieve secret data protected by SGX
	enclaves from it.

	\item \textbf{Memory Access Semantics}:
	CPU memory protection mechanisms physically block access to PRM from all
	external agents, by treating such accesses as references to non-existent
	memory. To access a page inside an enclave using MOV and other memory
	related instructions, the hardware checks the following:
	\begin{itemize}
		\item Logical processor is executing in ``enclave mode''.
		\item Page belongs to enclave that the logical processor is executing.
		\item Page accessed using the correct virtual address.
	\end{itemize}
	If any of these checks fails, the page access is treated as reference to
	nonexistent memory, or by signaling a fault. This guarantees that even a
	process with higher privilege levels will not be able to access enclave's
	memory.
\end{itemize}

\subsection{SGX Limitations}

\begin{figure}[ht]
\centering
\includegraphics[width=3.3in]{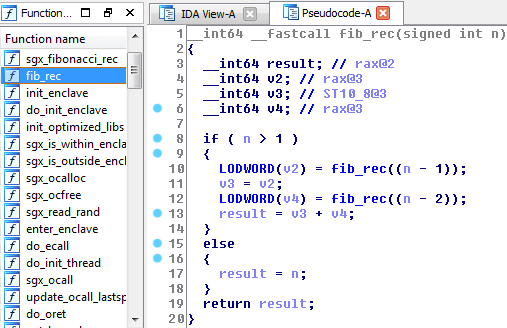}
\caption{Reversing engineering the enclave, a recursive Fibonacci function.}
\label{fig:fib_rev}
\end{figure}

As with most technologies, Intel SGX has some limitations that need to be
considered by developers. The main ones are the following:


\begin{itemize}
	\item \textbf{Code privacy}: The entire enclave code can be viewed from the
    executable file stored in the file system. The code is protected from 
    modification, but it does not allow
	developers to maintain their code private. There are many scenarios where
	developers want to keep their code private, so this poses as a serious
	privacy drawback, once attackers could reverse engineer the enclave code by
	disassembling it, or even generating a pseudocode very similar to the
  	original one, through tools like IDA\footnote{\url{https://www.hex-rays.com/products/ida/index.shtml}}. We performed this 
    attack and reverse-engineered an enclave (\textit{.so} file) containing a 
    recursive Fibonacci function, as depicted in Figure~\ref{fig:fib_rev}.

	\item \textbf{Static linking}: Applications that use third-party libraries
	need to statically link these libraries against their enclaves. This may
	result in generating a large footprint for enclaves and, consequently, waste
	space in memory -- a scarce resource for SGX.

	\item \textbf{Memory size}: When starting a machine, BIOS needs to reserve a
	portion of memory to the processor (PRM). Also, the	entire EPC must reside
	inside the PRM. In the current version of SGX, this portion of memory is
	limited to only $128\ MB$ in size per machine. If the space	needed is more
	than the space available, a large overhead in processing time is added, due to
	the need to encrypt the data before swapping from EPC to DRAM and decrypt
	the	data after swapping from DRAM to EPC. In~\cite{silva2017} experiments
	are presented, showing that randomly accessing memory is more than $100$
	times more costly in SGX enclaves that need $128\ MB$, which results in
	exceeding the PRM size, in comparison with unprotected applications.

\end{itemize}

\subsection{Vulnerabilities}
\label{sec:sgxvulnerabilities}

SGX protects against many types of attacks, even from privileged users and softwares. However, a
side-channel adversary is able to gather statistics from the CPU regarding execution and may be
able to use them to deduce characteristics of the software being executed.
Examples of analyses are power statistics, performance statistics including platform cache
misses, branch statistics via timing, and information on pages accessed via page tables. It is well
documented that SGX does not defend against side-channel adversaries~\cite{yuanzhong, brasser}.

SGX components are complex and unlikely to be bug-free, like any other software. There are drivers,
libraries, dependencies, and complex instructions available for developers. Moreover, enclave
developers may make mistakes and even the so-called protected areas may contain common
vulnerabilities like stack-based buffer overflows and uncontrolled format strings. This problem is boosted
by the limited portion of memory because it affects the effectiveness of the Address Space Layout
Randomization (ASLR).

ASLR is a security technique involved in protection from many types of attacks. In order to
prevent an attacker from reliably jumping to, for example, a particular exploited function in
memory, ASLR randomly arranges the address space positions of key data areas of a process,
including the base of the executable and the positions of the stack, heap and libraries. With SGX,
because the memory space for an enclave is quite small, a simple brute force mechanism can easily
identify the correct address. In our experiments, we observed that for different executions, an
element has its addresses changed by only two bytes, meaning that the randomization is for
approximately only $65536$ possibilities. This is very small, considering that an attacker can, for
example, increase the attack success probability through the injection of NOP slides before a
malicious code.

\section{DYNSGX: DYNAMICALLY LOADING FUNCTIONS INTO SGX ENCLAVES WITH PRIVACY
GUARANTEES}
\label{sec:dynsgx}

DynSGX aims at providing an alternative to the conventional SGX programming
model, where the entire code of the application to be run inside an enclave
needs to be included in or linked against the enclave at build time. Doing so
could cause the enclave size to rapidly exceed the available memory of the EPC
and, consequently, cause a large overhead in processing time due to the need to swap
pages. The conventional SGX programming model would also cause the
application code that runs in the enclave to be completely visible after loading
the enclave into memory.

Instead, DynSGX toolset allows users to start their secure application with a
very small enclave, and dynamically load functions into and remove functions
from the enclave at runtime as needed. This allows users and developers to
better manage the amount of memory that is occupied by the application
code. Also, by using the RA process to establish a secure communication channel
and using SGX capabilities to protect memory, DynSGX provides privacy guarantees for
the code loaded to the enclave at runtime.

\subsection{DynSGX TCB}
In DynSGX we limit the trusted computing base (TCB) to the Intel SGX SDK/PSW
plus a set of only six SGX-compatible C/C++ functions, yielding an enclave with
an initial size of only $1.4$MB. Such functions are used for $(i)$ performing
the SGX RA process, $(ii)$ loading functions into the enclave, $(iii)$ running
loaded functions inside the enclave and $(iv)$ unloading functions from the
enclave.

For the RA process, only the \textit{enclave\_ra\_init} function needs to be
placed inside the enclave. This function internally calls the
\textit{sgx\_ra\_init} function from the SGX SDK, which starts the RA process.
To load functions into the enclave, two functions are provided and placed inside
the enclave: \textit{enclave\_get\_fas}, which is responsible for providing
a list of functions that are already registered inside the enclave, and
\textit{enclave\_register\_function}, which is responsible for loading new
functions into the enclave. The \textit{enclave\_execute\_function} can be used
to securely execute the functions that were loaded into the enclave. Finally,
DynSGX provides the functions \textit{enclave\_unregister\_function} and
\textit{enclave\_clear\_functions} that developers can use to unload functions
from the enclave.

\subsection{DynSGX Programming Model}

DynSGX, as many other cloud-based tools, follows the client-server model. The
DynSGX enclave runs in the server side, and developers interact with it from the
client side.

DynSGX does not require developers to know how to develop SGX applications.
Instead, developers can write their functions as they were writing regular C
programs. After writing their functions, the tool compiles the $.c$ file that 
contains the functions and then retrieves the bytes that compose the compiled 
functions (this step can be done by using a tool called \texttt{bytes\_extractor}
provided as part of DynSGX). The bytes of the functions can then be sent to be 
loaded into the enclave, and later on be executed inside it. An important note 
is that the $.c$ files are compiled with the \emph{-fPIC} flag so that the compiled
code is position independent.

Let us consider an example where a user wants to securely process a
function to sum two integer numbers. This \textit{sum\_function} is depicted in
Listing \ref{lst:sumfunctionC}.

\noindent\begin{minipage}{.24\textwidth}
\begin{lstlisting}[language=C, label=lst:sumfunctionC, caption={Sample C function
for summing two integer numbers.}]
int sum(int a, int b) {
  return a + b;
}
\end{lstlisting}

\end{minipage}\hfill
\begin{minipage}{.21\textwidth}

\begin{lstlisting}[label=lst:sumfunctionASM, caption={Corresponding assembly for the
\textit{sum\_function}.}]
push %rbp
mov  %rsp,%rbp
mov  %edi,-0x4(%rbp)
mov  %esi,-0x8(%rbp)
mov  -0x4(%rbp),%edx
mov  -0x8(%rbp),%eax
add  %edx,%eax
pop  %rbp
retq
\end{lstlisting}
\end{minipage}

After compiling the file containing this function, the \texttt{bytes\_extractor}
tool extracts the function bytes from the assembled x86-64 code (Listing \ref{lst:sumfunctionASM}),
resulting in the following \texttt
{hexstring}: {\ttfamily \textbackslash x55\textbackslash x48\textbackslash x89\textbackslash
xe5\textbackslash x89\textbackslash x7d\textbackslash x
fc\textbackslash x89\textbackslash x75\textbackslash xf8\textbackslash
x8b\textbackslash x55\textbackslash xfc\textbackslash x8b\textbackslash x45\textbackslash
xf8\textbackslash x01
\textbackslash xd0\textbackslash x5d\textbackslash xc3}.
This \textit{hexstring} can be loaded into the DynSGX enclave, where it will
be registered and stored in the heap, a protected memory area. When the
user needs to execute this function, it will be casted to a regular
function. The user can unload their function from the enclave when it
is no longer needed, in order to free memory.

\subsection{Application Lifecycle}
DynSGX enclaves are started with only a limited number of essential functions
inside it. After the enclave is loaded, developers/users can contact it to
dynamically provision their functions. After sending the
functions, users can execute them inside the DynSGX enclave, and even
unload them afterwards. The steps needed to complete this process are
illustrated in Figure~\ref{fig:softwareLifecycle} and described as follows:

\begin{figure}
\centering
\includegraphics[width=3.3in]{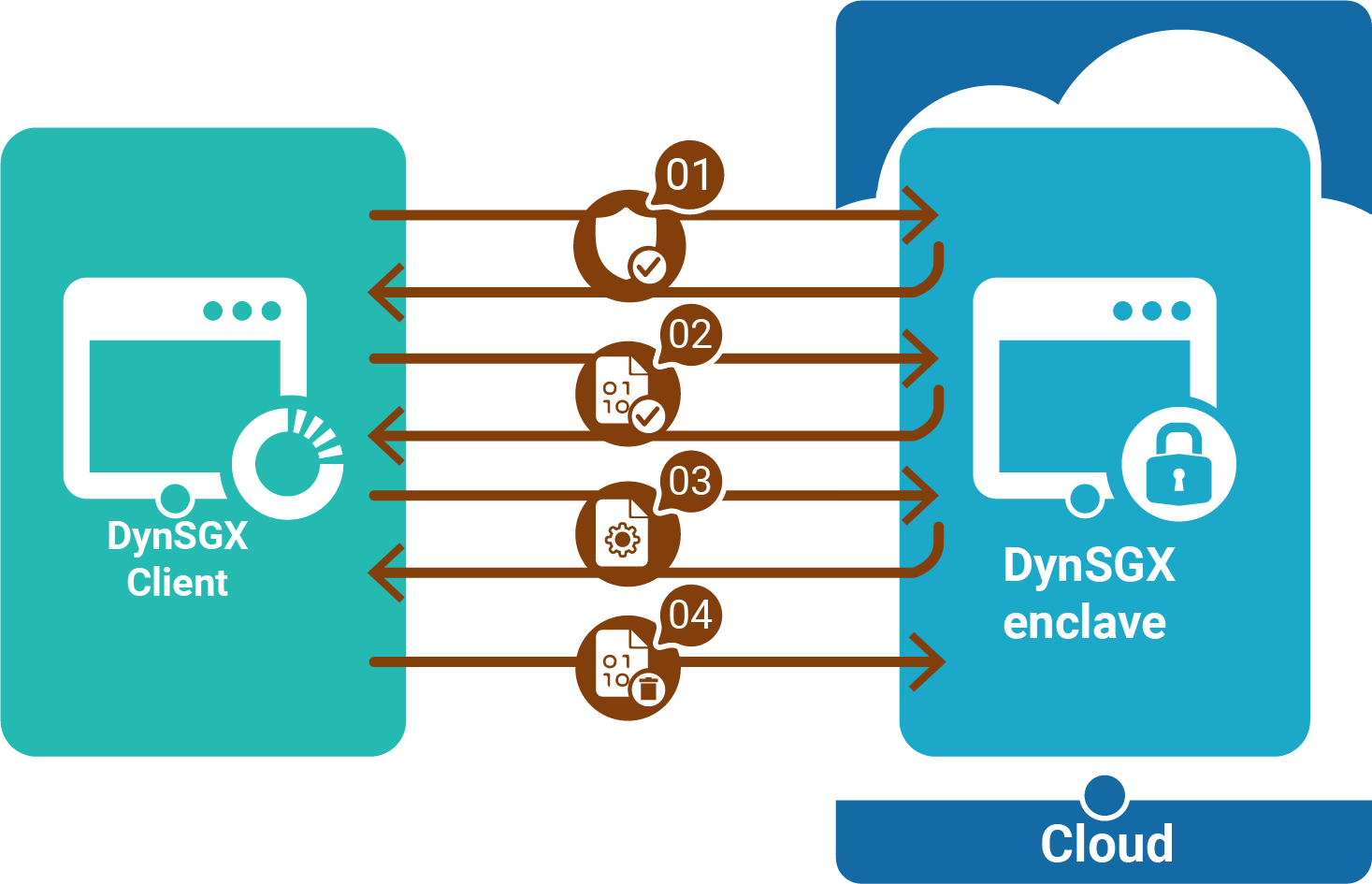}
\caption{DynSGX Software Lifecycle.}
\label{fig:softwareLifecycle}
\end{figure}

\begin{enumerate}
\item The client communicates with the DynSGX enclave and performs the RA
process to verify the identity of the enclave and establish a secure
communication channel between both;
\item The client compiles the user provided functions, extracts the bytes of the generated binary file using the \texttt{bytes\_extractor} tool, and sends them to the enclave via the secure communication channel. As a response, it obtains an identifier of the loaded function;
\item Via the secure channel, the client requests the DynSGX enclave to
execute a dynamically loaded function, along with any user provided function parameter. As a result, the client receives the result of the execution;
\item The user uses the client to request the DynSGX enclave to unload the
functions to free memory space.
\end{enumerate}

\subsection{Distributed Linking}
\label{sec:distributed_linking}

For a self-contained function (\textit{i.e.}, does not use external elements), compiling
and sending the bytes of the assembled code is enough. However, if the function
uses external elements, a distributed mechanism is needed  to map these elements
into their corresponding addresses at the enclave side. Examples of external
elements may be:

\begin{itemize}
  \item Library functions, such as from the \texttt{tlibc} (trusted libc, from the SGX
  SDK);
  \item Other functions previously defined by the user;
  \item Global variables.
\end{itemize}

DynSGX has an internal mechanism that retrieves the addresses and return types
of all functions available to the enclave (i.e.: \texttt{tlibc} functions, functions
loaded by the user, etc.) and sends to the client a JSON like the one presented
in Listing \ref{lst:jsonaddresses}, after the RA process is completed.

\begin{lstlisting}[language=C, label=lst:jsonaddresses, caption={JSON mapping the
external elements that can be used by a function into their corresponding addresses.}]
{
  "snprintf":  "(*(int(*)(0x7f1e438176f0)))",
  "vsnprintf": "(*(int(*)(0x7f1e4381d770)))",
  "strcmp":    "(*(int(*)(0x7f1e438179a0)))",
  ...
}
\end{lstlisting}

Therefore, at the client side, for a function like the one presented in Listing
\ref{lst:func_check_password}, the \textit{strcmp} at line 3 is replaced by
{\ttfamily (*(int(*)(0x7f1e438179a0)))}. This works because it is the same of
casting and calling a function pointer. The compiler does not know what will be
in this address, but in runtime, the \textit{strcmp} function will be at this
address within the enclave.

\begin{lstlisting}[language=C, label=lst:func_check_password, caption={Example
of a function that uses an external element (the strcmp function).}]
int check_password(char* input) {
  char password[] = "topsecret123";
  return !strcmp(input, password);
}
\end{lstlisting}

\subsection{Requirements}
To run the DynSGX enclave and load functions into it in runtime, three
requirements need to be met:

\begin{itemize}
\item \textbf{SGX-capable hardware}: SGX technology must be available and
enabled by BIOS. Such hardware is commercially available since the end of 2015.

\item \textbf{SGX driver}: SGX driver must be installed in
order to enable the OS and other softwares to access the SGX hardware.

\item \textbf{SGX PSW}: SGX PSW\footnoteref{fn:SGXSDK} is used to launch SGX
enclaves, and also to generate data necessary for the RA process. DynSGX
requires a small modification in the regular PSW. This modification regards the
option to make the program heap executable and is done by applying a patch to
the SGX PSW code\footnote{\url{https://patch-diff.githubusercontent.com/raw/01org/linux-sgx/pull/63.patch}}.
\end{itemize}

\subsection{Vulnerabilities}
\label{sec:vulnerabilitiesDynSGX}

Apart from side channel attacks and the vulnerabilities that an enclave's code may have, DynSGX
introduces a new attack surface: the function sent by the user. To provide its features,
DynSGX disables two security protections: stack canaries at the user's functions, and non-executable heap.

Stack canaries are used to detect a stack buffer overflow before execution of malicious code can
occur. This method works by placing a small integer in the memory, the value of which is randomly chosen at
program start, just before the stack return pointer. Most buffer overflows overwrite
memory from lower to higher memory addresses in order to overwrite the return pointer (and thus
take control of the process). Therefore, the canary value is also overwritten. This value is checked to
make sure it has not changed before a routine uses the return pointer on the stack. If the value
changes, the function \textit{\_stack\_chk\_fail} from \texttt{libc} is called. This function and its 
call, however, is introduced by the compiler. A C programmer is not able to access it, neither in the
enclave side to get its address, nor in the client side to replace its calling form. Therefore, the 
technique described in Section \ref{sec:distributed_linking} should not work for stack canaries.

Non-executable heap is a security protection that helps to prevent certain exploits from succeeding,
particularly those that inject and execute code in the heap. With DynSGX, user's functions 
are dynamically loaded into the heap space. Therefore, it was necessary to disable this protection using the configuration 
\texttt{<HeapExecutable>1</HeapExecutable>} at the \texttt{Enclave.config.xml} file.

Given the limitation of the ASLR mechanism due the small memory space, and the disabling of stack 
canaries and non-executable heap protections, it is very important to put more effort into 
developing secure codes. Listing \ref{lst:func_check_password_vuln1} presents an example of a 
vulnerable code that could be dynamically loaded using DynSGX.

\begin{lstlisting}[language=C, label=lst:func_check_password_vuln1, caption={Example of a function 
vulnerable to stack buffer overflow.}]
void check_password(char *input) {
    char buffer[16];
    char password[] = "topsecret123";
    strncpy(buffer, input, strlen(input));
    if (!strcmp(buffer, password))
        access();
}
\end{lstlisting}

This function is vulnerable to stack buffer overflow. If the attacker provides an input longer than 
15 bytes, it will overwrite other values that could be in the memory, like the value in the stack 
base pointer and the return address. Considering that the address of \textit{access} is \textit{0x7ffff580160d}, an attacker could have access overwriting the return 
address using the following payload: \textit{AAAAAAAAAAAAAAAAAAAAAAAA\textbackslash x0d
\textbackslash x16\textbackslash x80\textbackslash xf5\textbackslash xff\textbackslash
x7f\textbackslash x00\textbackslash x00}.

Another example of vulnerability is uncontrolled format string. A format function is a special kind
of C function that takes a variable number of arguments, from which one is the so called 
format string. If an attacker is able to provide the format string to a C format
function in part or as a whole, a format string vulnerability is present. By doing so, the attacker
can read and write to anywhere. Listing \ref{lst:func_check_password_vuln2} presents an example of
a code with uncontrolled format string vulnerability.

\begin{lstlisting}[language=C, label=lst:func_check_password_vuln2, caption={Example of a function 
vulnerable to uncontrolled format string.}]
void check_password(char *input) {
    char password[] = "topsecret123";
    if (!strcmp(input, password)) {
        access();
    } else {
        char error[30];
        snprintf(error, 30, input);
        strncat(error, " is incorrect!", 14);
        log_msg(error);
    }
}
\end{lstlisting}

In a normal situation, the effect of this function is the same of the one presented in Listing
\ref{lst:func_check_password}, but with the additional feature of logging an error message. The 
uncontrolled format string vulnerability is at line 7, and an attacker can provide the following
payload: {\ttfamily \%10\$p \%11\$p}. Therefore, the function will log the following message: {\ttfamily
0x6572636573706f74 0x33323174 is incorrect!}. The numbers in hexadecimal are the password 
representation in little endian format.

Given the limitations in the security protections, it is very important to write the code carefully
and do regular code reviews. The usage of tools that examine source code and report possible
vulnerabilities (usually sorted by risk level) may be useful. Flawfinder\footnote{\url{https://www.dwheeler.com/flawfinder/}},
Cppcheck\footnote{\url{http://cppcheck.sourceforge.net/}}
and CheckConfigMX \cite{braz} are examples of static analysis tools
that can be integrated with DynSGX for quickly finding and removing at least some potential
security problems before sending a code to the enclave.

\section{EVALUATION}
\label{sec:evaluation}

\begin{figure*}
\centering
\subfloat[Latency of \textit{sum\_array} function]
{\includegraphics[width=3.45in]{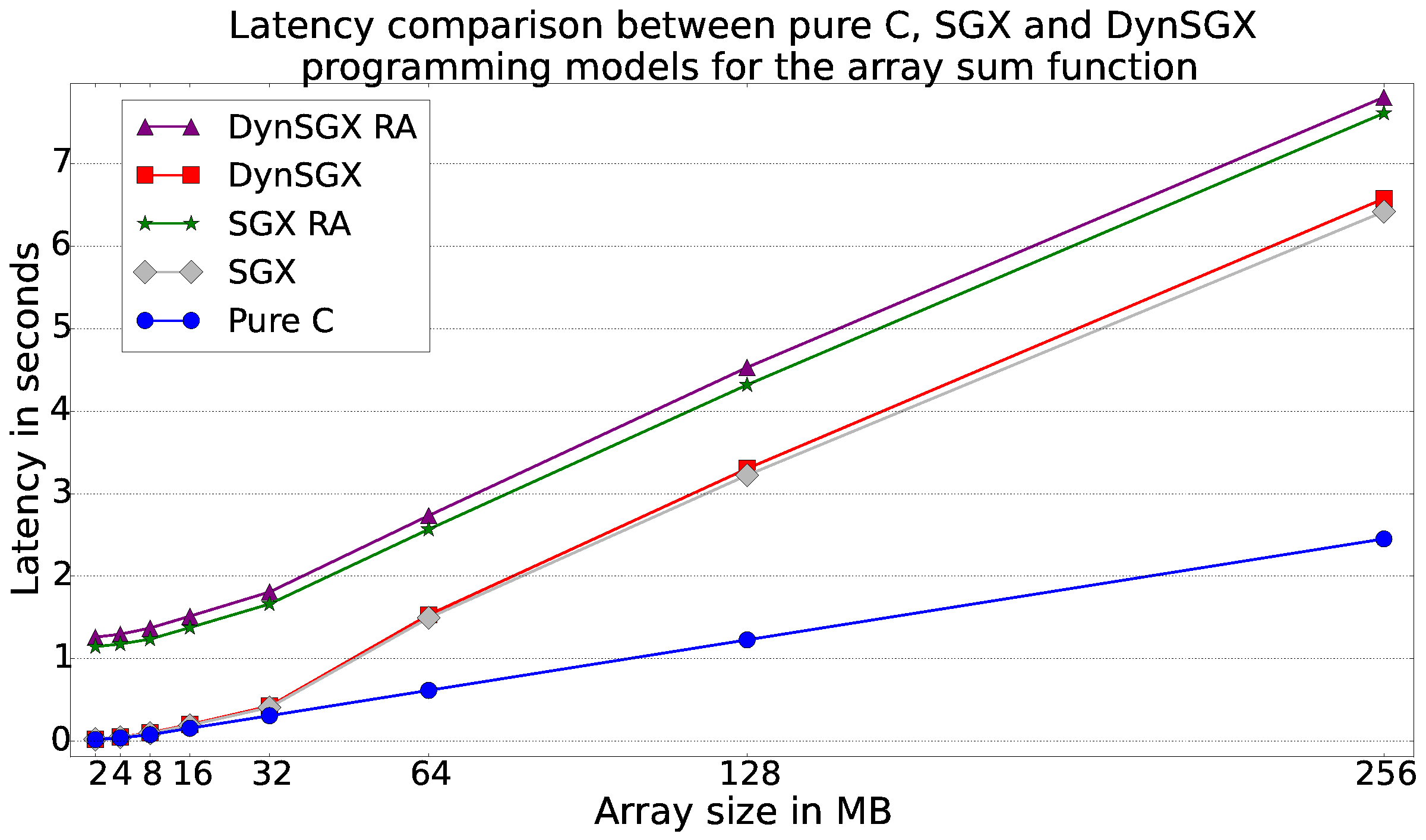}
\label{fig:sum_results}}
\hfil
\subfloat[Latency of \textit{recursive\_fibonacci} function]
{\includegraphics[width=3.45in]{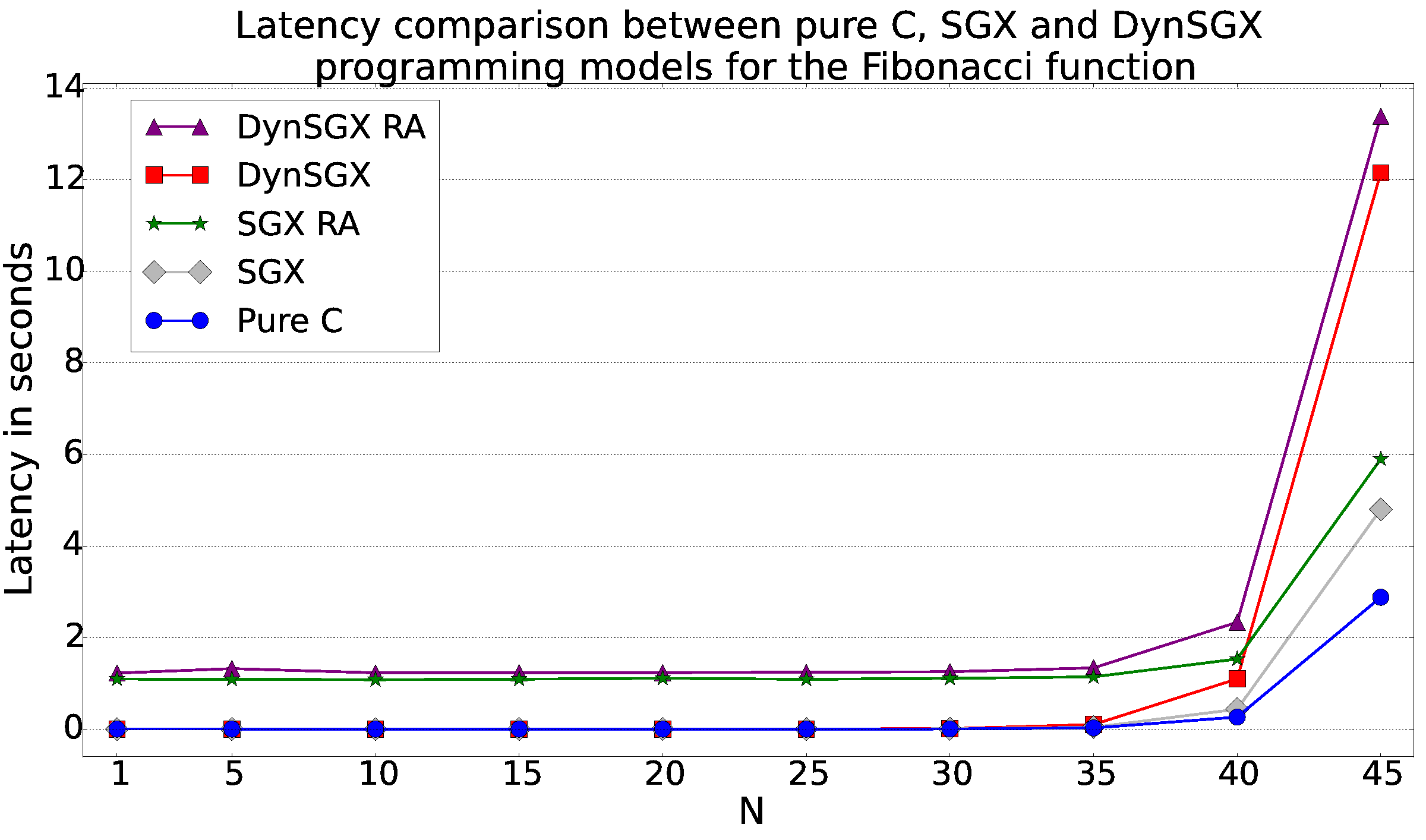}
\label{fig:fib_results}}
\caption{Latency comparison between pure C, Intel SGX and DynSGX implementations}
\label{fig:exp_results}
\end{figure*}

To assess the performance of applications that use the DynSGX toolset, we
performed a series of experiments that compare it to the pure C and regular SGX and programming models. In this section we describe the experiments performed, and present a
discussion based on the results obtained.

\subsection{Experiments Setup}

The experiments were conducted in a machine running Ubuntu Linux 16.04 with one
Intel i7-6700 SGX capable processor and $8\ GB$ of RAM. The server side of the
application is written in C++ and the client side is written in Python 2.7.

In our experiments, we measured the latency that represents the total time taken
from the client requesting the server to process a function until the client
receives the response from the server. The latency also includes the time taken
to perform the RA process when RA is needed, and to encrypt and decrypt data
both in the client and in the server to ensure privacy and security of data
exchanged between client and server. The latency was calculated based on two
different functions. The first one is the \textit{sum\_array} function, which
iterates over an array of integers and computes the sum of all of them, and
the second one is the \textit{recursive\_fibonacci} function, which recursively
calculates the $n$-th term on a Fibonacci sequence. These two functions were selected to illustrate a memory-intensive application (a known limitation of SGX) and CPU-intensive application.

Three different
implementations were made. The first does not consider
 any security and privacy guarantees. The second uses
 regular SGX programming model, with privacy and security guarantees of
the data being processed but no guarantees regarding the privacy of the code.
The third uses the DynSGX programming model, which
enables better manageability of the memory consumed by enclave functions and
privacy of the code being executed inside the enclave. For the SGX and
DynSGX implementations, we considered both the cases where the RA process is
needed to establish a secure communication channel, and where the secure
communication channel had already been established.

\subsection{The Experiments}
The experiments consisted of two parts. The first part aimed to compare the
latency for the computation of the \textit{sum\_array} function for arrays with
$2$, $4$, $8$, $16$, $32$, $64$, $128$, and $256$ MB in size. Figure~\ref
{fig:sum_results} depicts the median values of latency obtained in this part. This
experiment is useful to understand the behavior of using DynSGX with iterative
functions and with higher memory usage.

The second part aimed to compare the latency for the computation of the \textit
{recursive\_fibonacci} for $n$ values of $1$, $5$, $10$, $15$, $20$, $25$, $30$,
$35$, $40$, and $45$. Figure \ref{fig:fib_results} depicts the median
values of latency in this part. This experiment is important to understand the
behavior of using DynSGX with recursive functions.

In both parts of the experiment, for each array size and $n$ value, 30 runs
were executed.

\subsection{Discussion}

As can be seen in Figure \ref{fig:exp_results}, both SGX and DynSGX
implementations always exhibit considerable overheads when compared to the
unprotected pure-C implementation. This is due to two main factors: $(i)$ the
time taken to perform the RA process and $(ii)$ the need to encrypt and decrypt
data.

From the experiments results we can also observe that for processing iterative
functions, DynSGX has only a small overhead compared to the regular SGX
use. This behavior can be observed both when the RA process is needed and when
a secure channel has already been established. In this case, we consider the benefits
of using DynSGX significant because of the code privacy guarantees and
the possibility to manage memory occupied by code.

Nevertheless, it is important to notice that DynSGX may not perform as
well when processing recursive functions. Figure \ref{fig:fib_results} depicts this situation, where the \textit
{recursive\_fibonacci} function grows exponentially in number of recursive
calls. In this case, DynSGX performs much slower than the regular SGX
implementation. This is due to the fact that with DynSGX the function code
resides in the heap, which is considered a data segment, and competes with other data
areas (such as the stack frames of each recursive call) for a space in the processor
cache. Modern processors have different caches for instructions and data, hence, the
advantage of regular SGX with code at the instructions segment.
This behavior is not limited to DynSGX; a pure C or regular SGX implementation of a recursive algorithm that stores code in the heap also suffers some overhead. 

Regarding security aspects, it is also important to note that disabling stack canaries and enabling heap execution may not imply a risk to the hardware owner, since in a cloud environment it is possible to set up other security layers. For example, the DynSGX enclave could run inside an isolated environment of a virtual machine.

Table~\ref{tab:comparison} contains a comparison of advantages and
disadvantages of each of the three implementations considered in our experiments.

\begin{center}
    \begin{table}
        \caption{Performance Comparison Between Pure C, SGX and DynSGX Implementations}
        \label{tab:comparison}
        \begin{tabular}{|c|c|c|c|c|c|c|}
            \cline{2-7}
            \multicolumn{1}{ c |}{} & \multicolumn{2}{ c |}{\textbf{Integrity}} & \multicolumn{2}{ c| }{\textbf{Privacy}} & \multicolumn{2}{ c |}{\textbf{Performance}} \\ \cline{1-7}
            \textbf{Impl.} & \textbf{data} & \textbf{code} & \textbf{data} & \textbf{code} & \textbf{iterative} & \textbf{recursive} \\ \hline
            Pure C &  &  &  &  & High & High \\ \hline
            SGX & \ding{51} & \ding{51} & \ding{51} &  & Medium & Medium \\ \hline
            DynSGX & \ding{51} & \ding{51} & \ding{51} & \ding{51} & Medium & Low \\ \hline
        \end{tabular}
    \end{table}
\end{center}
\section{RELATED WORK}
\label{sec:relatedwork}
Intel has recently introduced SGX2, which extends the SGX instruction set by adding
support for dynamic memory management from inside SGX enclaves \cite
{mckeen2016intel}. SGX2 will allow lazy loading enclave code into the EPC,
instead of loading it at once at enclave load time. This will avoid the need of
DynSGX to make the modification to the SGX PSW described on Section~\ref
{sec:dynsgx}. On the other hand, SGX2 does not address any of the code privacy
concerns addressed by DynSGX. Besides that, SGX2 does not aim at enabling
developers to create new functions and loading them into enclaves after they
have been built, but only to dynamically load and unload code that is already
linked against enclaves.

In \cite{arnautov2016scone} SCONE is proposed. SCONE enables developers to
compile their C applications into Docker containers protected with the SGX
capabilities. Developers can simply compile their C applications with a special
compiler that is part of the SCONE toolset, and deploy the compiled application
in a SCONE client, and it will transparently be protected by SGX capablities.

SecureWorker\footnote{\url{https://www.npmjs.com/package/secureworker}} is an NPM package that allows JavaScript code
to be run inside SGX enclaves. It is still under development, and many important
features (\textit{e.g.}, remote attestation) are yet to be implemented.

On the one hand, both SCONE and SecureWorker solutions, like DynSGX, ease the development
of SGX solutions. On the other hand, both SCONE and SecureWorker lack the
capabilities of dynamically loading code into SGX enclaves and preserving the
privacy of such code.

A known approach for code privacy is obfuscation \cite{linn}. It makes the code difficult
for humans to understand and programmers may deliberately obfuscate 
code to conceal its purpose (security through obscurity), primarily, in order to 
prevent reverse engineering. However, Barak \textit{et al.} \cite{barak} perform a 
theoretical investigation and prove that, secure obfuscation is impossible. It is 
known that a talented hacker can reverse engineer code even after being obfuscated.
\section{CONCLUSION}
\label{sec:conclusion}

Intel SGX has been considered to be one of the most promising TEE technologies. TEEs can enable application with sensitive data to run on the cloud.
However, its limitations and code privacy concerns have drawn its applicability
in cloud environments into question. This paper presented a toolset
that enables users to dynamically deploy their functions into enclaves and also ensures the privacy of such functions.
Our evaluation shows that
DynSGX enables developers to load/unload functions into/from enclaves and can be
used to guarantee the privacy of code to be executed inside SGX enclaves with a
small overhead compared to the regular SGX programming model. We have also
shown that recursive functions with many recursive calls may not be suitable for
use with DynSGX.

As future work we plan to extend DynSGX to provide support for
programming languages other than C. In addition, we want to provide the developer/user with functions that enable a more detailed monitoring and, consequently, a more efficient management 
of the enclave memory.

\section*{Acknowledgements}

This research was partially funded by EU-BRA SecureCloud project  (EC, MCTIC/RNP, and SERI, 3rd Coordinated Call, H2020-ICT-2015
Grant agreement no. 690111) and by CNPq, Brazil.



\bibliographystyle{IEEEtran}

\end{document}